\documentclass[
prl,
reprint,
twocolumn,
floatfix
]
{revtex4-1}

\usepackage{float}

\usepackage{latexsym}
\usepackage{amssymb, amsmath}
\usepackage{exscale}
\usepackage{bm, graphics}
\usepackage{graphicx, subfigure, pstricks}

\newcommand{\D}{\mbox{\rm d}}
\newcommand{\uhb}[1]{\underline{\hat{\bf{#1}}}}
\newcommand{\hb}[1]{\hat{\bf{#1}}}

\begin{document}
\title{Cavity-assisted spontaneous emission
of a single three-level atom
as a single-photon source
}

\author{M. Khanbekyan}
\email[E-mail address: ]{mkh@tpi.uni-jena.de}

\author{D.-G. Welsch}
\affiliation{Theoretisch-Physikalisches Institut,
Friedrich-Schiller-Universit\"at Jena, Max-Wien-Platz 1,
D-07743 Jena, Germany}

\date{\today}

\begin{abstract}
Stimulated Raman interaction of a
classically pumped
single three-level $\Lambda$-type
atom in a resonator cavity featuring
both radiative and unwanted losses
is studied. It is shown that
in the regime of
stimulated adiabatic Raman passage
the excited outgoing wave packet
of the cavity-assisted
electromagnetic field can be prepared in a
one-photon Fock state
with high efficiency.
In this regime, the spatio-temporal
shape of the wave packet
does not depend on the
interaction shape of the pump field, provided that the
interaction of the atom with the pump
is time-delayed with respect to the
interaction of the atom with the
cavity-assisted field.
Therefore, the scheme can be used
to generate a sequence of
identical radiation pulses
each
of almost one-photon Fock
state.
It is further shown that the
spatio-temporal shape of the outgoing wave packet
can be manipulated by
controlling the time of interaction between the atom and
the cavity-assisted electromagnetic field.
\end{abstract}


\maketitle

\section{Introduction}
\label{introduction}

The interaction of a single atom with a
quantized radiation-field mode in a high-$Q$ cavity
serves as a basic ingredient in various
schemes in quantum information
science (for a review see, e.\,g.,
Refs.~\cite{walther:617, walther:1325}).
In this context, quantum control of single-photon
emission from an atom in a cavity
for generating one-photon Fock states on demand
has been an essential prerequisite~\cite{monroe:238}.
Single-photon sources operating
on the basis of vacuum-simulated adiabatic passage with just
a single atom in a \mbox{high-$Q$} optical cavity has been realized,
where adjustment of the spatio-temporal shapes
of the outgoing wave packets associated with the emitted photons
has been achieved
by means of driving laser pulses~\cite{kuhn:067901, keller:1075}.

More recently, we have discussed
the possibility to
use a single two-level atom in a high-$Q$ cavity as a
single-photon emitter, with the wave packet
associated with the emitted photon being
shorter than the cavity decay time and its spatio-temporal shape
being time-sym\-\mbox{metric~\cite{khanbekyan:013822,fidio:043822}}.

In the following we consider,
within the frame of exact quantum electrodynamics,
the interaction of a pumped three-level
$\Lambda$-type atom with a realistic cavity-assisted electromagnetic
field, with the aim of one-photon Fock-state emission.
In particular, we study in detail both the properties of the
excited outgoing wave packet
and the quantum state it is prepared in.
Further, the general explicit expression obtained for the spatio-temporal shape
of the outgoing wave packet allows us to study different
coupling
regimes of atom--pump and atom--cavity-field interactions.


\section{Basic equations}
\label{sec3.2}

We consider an atom (position ${\bf r}_A$)
that interacts with the electromagnetic field
in the presence of a dispersing and absorbing dielectric medium.
Applying the multipolar-coupling scheme in electric dipole
approximation, we may write the Hamiltonian
that governs the temporal evolution
of the overall system, which consists of the electromagnetic
field, the dielectric medium (including the dissipative degrees
of freedom), and the atom coupled to the field,
in the form of~\cite{knoell:1, vogel}
\begin{align}
   \label{1.1}
        \hat{H} =
&
        \int\! \D^3{r} \int_0^\infty\! \D\omega
      \,\hbar\omega\,\hb {f}^{\dagger}({\bf r },\omega)\cdot
      \hb{ f}({\bf r},\omega)
\nonumber\\&
+
	  \sum _k
        \hbar
\omega _{k}
\hat{S} _{kk}
-
   \hb{ d}_A\cdot
        \hb{E}({\bf r}_A)
.
\end{align}
In this equation, the first term is the Hamiltonian of
the \mbox{field--me}\-dium system, where the
fundamental
bosonic fields
\mbox{$\hb{ f}({\bf r},\omega)$}
and \mbox{$\hb{f}^\dagger({\bf r},\omega)$},
\begin{align}
    \label{1.3}
&      \bigl[\hat{f}_{\mu} ({\bf r}, \omega),
      \hat{f}_{\mu'} ^{\dagger } ({\bf r }',  \omega ') \bigr]
      = \delta _{\mu \mu'}\delta (\omega - \omega  ')
      \delta ^{(3)}({\bf r} - {\bf r }') ,
\\
\label{1.3-1}
&\bigl[\hat{f}_{\mu} ({\bf r}, \omega),
      \hat{f}_{\mu'} ({\bf r }',  \omega ') \bigr]
= 0,
\end{align}
play the role of the canonically conjugate system variables.
The second term is the Hamiltonian of the atoms, where
the $\hat{S}_{Ak'k}$ are the atomic flip operators for
the atom,
\begin{equation}
   \label{1.5}
   \hat{S} _{Ak'k} =
   | k'\rangle_{\!A} {_A}\!\langle k |
,
\end{equation}
with the $|k\rangle_{\!A}$ being the energy eigenstates
of the atom. Finally, the last term is the atom--field
coupling energy, where
\begin{equation}
   \label{1.7}
    \hb{ d}_A = \sum _{kk'}
    {\bf d} _{Akk'}  \hat{S} _{Akk'}
\end{equation}
is the electric dipole-moment operator of the atom
($ {\bf d} _{Akk'}$ $\!=$ $\!{_A}\!\langle k|
\hb{ d}_{\!A} | k' \rangle_A$), and the operator of the
medium-assisted electric field $\hb{E}({\bf r})$
can be expressed in terms of the variables
$\hat{\mathbf{f}}(\mathbf{r},\omega)$ and
$\hat{\mathbf{f}}^\dagger(\mathbf{r},\omega)$ as
follows:
\begin{equation}
\label{1.9}
\hb{E}({\bf r}) = \hb{E}^{(+)}({\bf r})
        +\hb{E}^{(-)}({\bf r}),
\end{equation}
\begin{equation}
\label{1.10}
\hb{E}^{(+)}({\bf r}) = \int_0^\infty \D\omega\,
      \uhb{E}({\bf r},\omega),
\quad
\hb{E}^{(-)}({\bf r}) =
[\hb{E}^{(+)}({\bf r})]^\dagger,
\end{equation}
\begin{multline}
      \label{1.11}
      \uhb{ E}({\bf r},\omega) =
\\
\nonumber\\
i \sqrt{\frac {\hbar}{\varepsilon_0\pi}}\,
\frac{ \omega^2}{c^2}
      \int \D^3r'\sqrt{\varepsilon''({\bf r}',\omega)}\,
      \mathsf{G}({\bf r},{\bf r}',\omega)
      \cdot\hb{f}({\bf r}',\omega),
\end{multline}
where the classical (retarded)
Green tensor $\mathsf{G}({\bf r},{\bf r}',\omega)$
is the solution to the equation
\begin{equation}
      \label{1.13}
      \bm{\nabla}
\times
    \bm{\nabla}\!  \times \mathsf{G}  ({\bf r }, {\bf r }', \omega)
      - \frac {\omega ^2 } {c^2} \,\varepsilon ( {\bf r } ,\omega)
      \mathsf{G}  ({\bf r}, {\bf r }', \omega)
      =  \bm{\delta} ^{(3)}  ({\bf r }-{\bf r }')
      \end{equation}
and satisfies the boundary condition at infinity, i.\,e.,
$\mathsf{G}({\bf r},{\bf r}',\omega)\to 0$ if
$|\mathbf{r}-\mathbf{r}'|\to\infty$.


\section{
Pumped three-level
$\Lambda$-type atom in a cavity}
\label{sec3.3}

\begin{figure}[t]
\includegraphics[width=.9\linewidth]{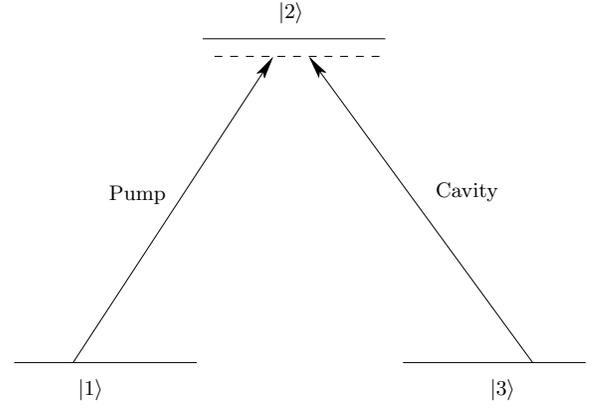}
\caption{\label{fig1}
Scheme of relevant energy levels and transitions for the three-level
$\Lambda$-type atom.
}
\end{figure}

Let us consider a three-level $\Lambda$-type atom
(Fig.~\ref{fig1}) in
a cavity bounded by a perfectly reflecting mirror
and a fractionally transparent (coupling) mirror and
model the cavity by
a one-dimensional dielectric
layer system (see, e.g., Ref.~\cite{khanbekyan:013822}.
In particular, we assume that the atomic
transition $|2\rangle\rightarrow|3\rangle$
is strongly coupled to the cavity field. Further,
restricting our attention to
\mbox{(quasi-)}resonant
atom--field interaction, we may start from the
(one-dimensional version of the) Hamiltonian~(\ref{1.1})
and apply the rotating-wave approximation to the
atom--field interaction.
Moreover, we assume that an external (classical) pump field with
frequency $\omega_p$ is applied to the
$|1\rangle\rightarrow|2\rangle$ transition of the atom.
In this way we may write the Hamiltonian in the form of
\begin{align}
   \label{1.15}
        \hat{H} =
&
        \int\! \D z\int_0^\infty\! \D\omega
      \,\hbar\omega\,\hat {f}^{\dagger}(z, \omega)
      \hat{f}(z, \omega)
+ \hbar \omega _{0} \hat {S}_{22}
\\&
  -g_c(t)\left[
        d_{23}\hat{S}_{32}^{\dagger}
            \hat{E}^{(+)}(z_A)
            +
\mbox{H.c.} \right]
\\[.5ex]&
-
\frac{\hbar}{2}
\Omega
_p(t)\left[
        \hat{S}_{12}^{\dagger}
            e^{i\omega_p t}
            +
\mbox{H.c.} \right].
\end{align}
Here, $\Omega_p(t)$
is the (real, time-dependent)
Rabi frequency of the pump field, and
the (real) function $g_c(t)$
defines the (time-dependent) shape of the
interaction of the atom with the cavity field,
which can be realized by (quasi-static) motion of the atom through the
cavity in the direction perpendicular to the cavity axis.

In what follows we assume that
the atom is initially (at time $t$ $\!=$ $\!0$)
prepared in the state $|1\rangle$ and the rest of
the system, i.\,e., the part of the
system that consists of the
electromagnetic field and the cavity, is prepared in the ground
state $|\{0\}\rangle$.
In order to avoid lengthy formulae
we assume that the frequencies
of the atomic transitions
$|1\rangle\leftrightarrow|2\rangle$ and
$|3\rangle\leftrightarrow|2\rangle$
are equal to each other
($\omega_{21}=\omega_{23}\equiv\omega_0$).
We may then expand the state
vector of the overall system at later times $t$ ($t\ge 0$) as
\begin{multline}
\label{1.17}
	|\psi(t)\rangle =
   C_1(t)|1\rangle|\!\left\lbrace0\right\rbrace\!\rangle+
  C_2(t)e^{-i\omega_{0} t}
  |2\rangle|\!\left\lbrace0\right\rbrace\!\rangle
\\[.5ex]
   +\int \D z\, \int_0^\infty \D \omega\,
   C_3(z, \omega, t) e^{-i\omega t}
   |1\rangle\hat{f}^{\dagger}(z, \omega)
   |\!\left\lbrace0\right\rbrace\!\rangle,
\end{multline}
where $\hat{f}^{\dagger}(z, \omega)|\{0\}\rangle$ is
an excited single-quantum state of the combined
field--cavity system.

It is not difficult to prove
that the Schr\"odinger equation for $|\psi(t)\rangle$
leads to the following system of
(in\-tegro-)differential
equations for the probability amplitudes
$C_1(t)$, $C_2(t)$ and $C_3(z, \omega, t)$:
\begin{equation}
  \label{1.19}
  \dot {C_1} =
      \frac{i}{2}
\Omega
_p(t)
      e^{i\Delta_pt}
      C_2(t),
\end{equation}
\begin{multline}
 \label{1.20}
 \dot {C_2} =
 \frac{i}{2}
\Omega
_p(t)
      e^{-i\Delta_pt}
      C_1(t)
\\[.5ex]
      -\frac{d_{23}}{\sqrt{\pi \hbar \varepsilon _0  \mathcal{A}}}
      \int_0^\infty\! \D\omega\, \frac{\omega ^2}{c^2}
      \int \D z
      \sqrt{\varepsilon''(z,\omega)}\,
\\[.5ex]
 \times
      G(z_A, z,\omega)
      C_3(z, \omega, t)
      e^{-i (\omega - \omega_0)t},
\end{multline}
\begin{multline}
  \label{1.21}
 \dot {C_3}(z, \omega, t) =
      \frac{d_{23}^*}{\sqrt{\pi \hbar \varepsilon _0  \mathcal{A}}}
      \frac{\omega ^2}{c^2}
       \sqrt{\varepsilon''(z,\omega)}\,
\\[.5ex]
\times
      G^*(z_A, z,\omega)
      C_2(t)
      e^{i (\omega - \omega_0)t},
\end{multline}
where $\mathcal{A}$ is the area
of the coupling mirror of the cavity,
and $\Delta_p$ is the detuning of
the pump frequency
from the atomic transition frequency
($\Delta_p=\omega_p-\omega_0$).
The spectral response of the cavity
field is known to be determined by the Green function $G(z,z',\omega)$. For a
sufficiently high-$Q$ cavity, the excitation spectrum effectively turns into
a quasi-discrete set of lines of mid-frequencies $\omega_k$
and widths $\Gamma_k$, according to the poles of the
Green function at the complex frequencies
\begin{equation}
  \label{1.30}
    \Omega _{k} = \omega_{k}
      - {\textstyle\frac {1} {2}}i\Gamma _{k},
\end{equation}
where the line widths are much smaller than the
line separations,
\begin{equation}
\label{1.30-1}
\Gamma_k \ll {\textstyle\frac{1}{2}}(\omega_{k+1} - \omega_{k-1}).
\end{equation}
In this case,
substituting the formal solutions to Eqs.~(\ref{1.19}) and (\ref{1.21})
[with the initial condition \mbox{$C_1(0)=1$}, \mbox{$C_2(0)=0$} and
\mbox{$C_3(z, \omega, 0)=0$}] into  Eq.~(\ref{1.20}), we
derive the integro-differential equation
\begin{align}
 \label{1.22}
 \dot {C_2} =
 \frac{i}{2}
\Omega
_p(t)
      e^{-i\Delta_pt}
 +
\int_0^t \! \D t'\,
    K(t,t')
    C_2(t'),
\end{align}
where the kernel function $K(t,t')$ reads
(for details, see Ref.~\cite{khanbekyan:013822})
\begin{align}
  \label{1.24}
  K(t,t')=
&
  -\frac{1}{4}
\Omega
_p(t)
\Omega
_p(t')
    e^{-i\Delta_p(t-t')}
\nonumber\\[1ex]&
    -
    \frac{1}{4}
   \alpha_k \Omega_k
   g_c(t)
    g_c(t')
     e^{-i(\Omega_k - \omega_0)(t-t')},
\end{align}
\begin{equation}
  \label{1.39}
   \alpha_k  =  \frac{4|d_{21}|^2 }
    {\hbar \varepsilon _0  \mathcal{A}|n_1
(\Omega_k)
    |^2 l}\,
    \sin^2
    [\omega_k |n_1
(\Omega_k)
    |
    z_A/c]
,
\end{equation}
where $l$ is the length of the cavity,
$n_1(\omega)$
is the (complex) refractive
index of the medium inside the cavity,
and $\omega _k$ is
now the fixed quasi-discrete line frequency that is
\mbox{(quasi-)}resonant
to the atomic transition frequency.

Following Ref.~\cite{khanbekyan:013822}, it can be shown that
when the Hilbert space of the system is
effectively spanned only by a single excitation,
on a time-scale that is short compared to the inverse spontaneous
emission rate of the atom the quantum state of the outgoing field
is given by means of the multimode Wigner function
\begin{equation}
\label{7.73}
W_{\mathrm{out}} (\alpha_i, t)
= W_1(\alpha_1,t)
\prod_{i\neq 1}
     W_i^{(0)}(\alpha _i, t),
\end{equation}
where
\begin{equation}
\label{7.74}
W_1(\alpha,t)
= [1-\eta(t)]W_1^{(0)}(\alpha)
     +\eta(t)W_1^{(1)}(\alpha),
\end{equation}
with $W_i^{(0)}(\alpha)$ and $W_i^{(1)}(\alpha)$
being the Wigner functions of the vacuum
state and the one-photon Fock state, respectively, for the
$i$th nonmonochromatic mode. As we see, the
mode labeled by the subscript \mbox{$i$ $\!=$
$\!1$},
i.\,e., the mode associated with the excited outgoing
wave packet,
is
basically
prepared
in a mixed state of a one-photon Fock state and the
vacuum state, due to
unavaoidable existence of unwanted losses.
The other nonmonochromatic modes of the
outgoing field with \mbox{$i$ $\!\neq$ $\!1$} are in the vacuum
state and, therefore, remain unexcited.
The Wigner function
$W_1(\alpha,t)$
reveals that
$\eta(t)$ can be regarded as being the efficiency
to prepare the excited outgoing wave packet in
a one-photon Fock state,
\begin{equation}
 \label{5.19}
         \eta(t)
=
\int_0^{\infty}\!
         \D\omega\, |F(\omega ,t)| ^2
\simeq
\int_{-\infty}^\infty
         \D\omega\, |F(\omega ,t)| ^2
,
 \end{equation}
where
\begin{multline}
  \label{7.1}
    F(\omega, t)=
    \frac{d_{21}}{\sqrt{\pi \hbar \varepsilon _0  \mathcal{A}}}
    \sqrt{\frac{c}{\omega}}
    \frac{\omega^2}{c^2}
\\[1ex]
\times
      \int ^t _0 \D t'\,
     G^*(0^+, z_A, \omega)
     C_2^*(t') e^{i\omega(t-t')}
     e^{i
\omega
_0t'}
.
\end{multline}

The excited outgoing wave packet is
characterized by the mode function
\begin{equation}
  \label{5.17}
    F_1(\omega,t)
    = \frac{F(\omega , t)}{\sqrt{\eta(t)}}\,
,
\end{equation}
and its
spatio-temporal shape
reads
\begin{multline}
  \label{1.51}
  \phi_1(z,t)=
  \frac{ \kappa _k}{2}
 \sqrt{
  {\displaystyle
        \frac{\pi\hbar\omega_k}
        {\varepsilon_0 c\mathcal{A} \eta(t)}
      }}
\\[.5ex] \times\!
     \int
     \D  t'\,
     g_c(t')
     C_2^*(t')
     e^{-i(\Omega_k^*- \omega_0) t'}
     e^{i \Omega_k^*(t-z/c)}
\\[1ex] \times\!
     [
  \Theta(z+l)
  \Theta(-z)
  \Theta(t-t')
  + \Theta(z)
  \Theta(t-t'-z/c)
  ]
  \end{multline}
(for details, see Ref.~\cite{khanbekyan:013822}).
Here,
the term with $\Theta(-z)$ corresponds to the part
of the excited outgoing mode that is still inside the cavity,
while the term with $\Theta(z)$ stands for the part which is
already escaped from the cavity.

\section{Results and Discussion}
\label{sec5}

Let us first consider the case when the atom interacts with the
pump field and the cavity-assisted electromagnetic field in the
"counter-intuitive" order,
where the atom--cavity-field interaction is well-established before
the pump field is turned on. Then, provided that the pump field
grows sufficiently slowly, the
evolution of the quantum state of the combined atom--cavity-field system
closely follows the one of a dark state,
i.\,e., the dressed state of
a single-mode--atom system
which has no contribution from the upper
atomic state $|2\rangle$.
Thus, stimulated Raman adiabatic passage
is realized, thereby a single photon
being generated in the (excited) outgoing field.

\begin{figure}[H]
\rput(7,4.5){a)}
\includegraphics[width=.95\linewidth]{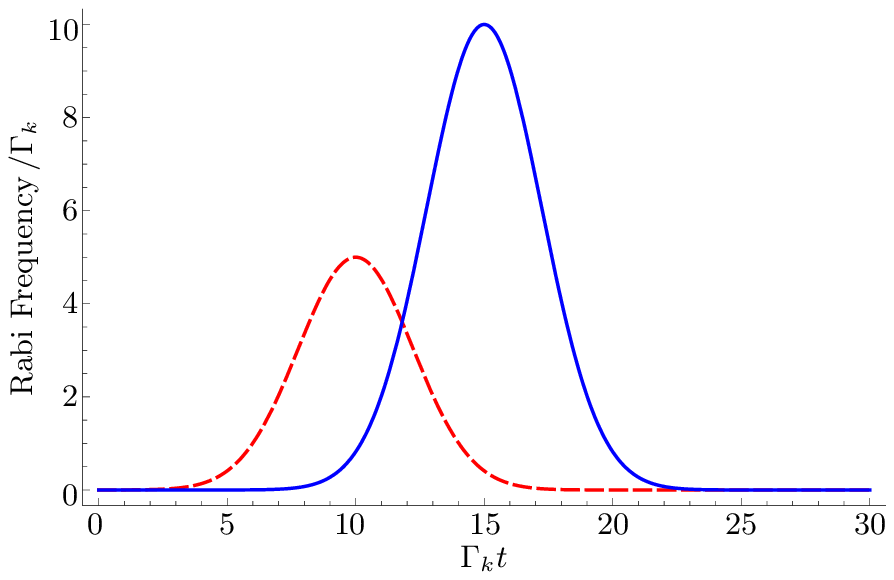}
\\[2ex]
\rput(7,4.5){b)}
\includegraphics[width=.95\linewidth]{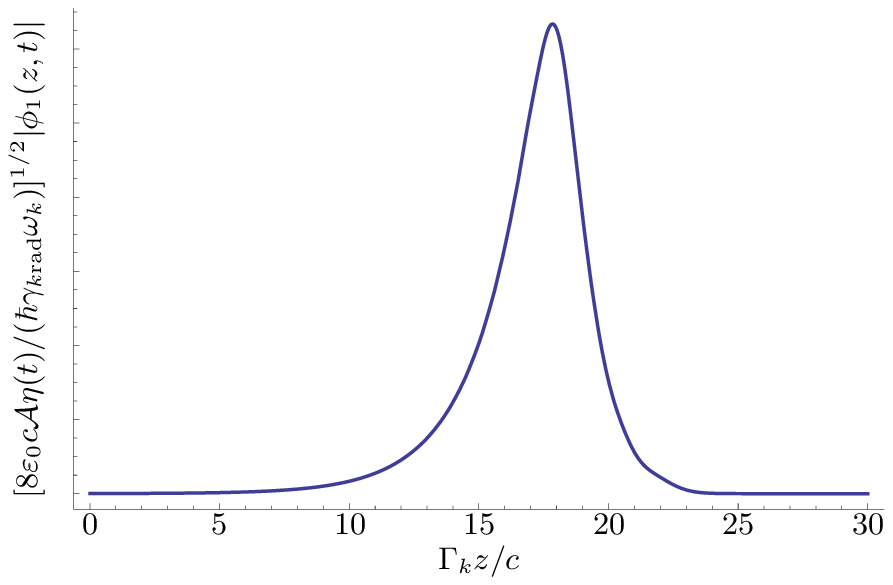}
\caption{\label{fig5.1}
(Color online)
a) Sequence of the interactions of the atom with the cavity field
[$R_k g_c(t)$, dashed curve] and
with the pump field [$\Omega _p(t)$, solid curve].
b) Numerical result for the spatio-temporal
shape of the excited outgoing mode for $\Gamma_k t$ $\!=$ $\!30$.
}
\end{figure}

For a numerical evaluation of the equations given above,
we have modeled the
(time-dependent) Rabi frequency
of the pump field, $\Omega_p(t)$,
and the (time-dependent) atom--cavity-field interaction
shape function $g_c(t)$ by
Gaussian functions
(see Fig.~\ref{fig5.1}),

\begin{equation}
  \label{5.1}
\Omega
_p(t) =
\Omega
_p \exp\left[-\frac{(t-t_p)^2}{\sigma_p^2}\right],
\end{equation}
\begin{equation}
  \label{5.3}
   g_c(t) = \exp\left[-\frac{(t-t_c)^2}{\sigma_c^2}\right].
\end{equation}
Further, we have assumed that, apart from the atom, the cavity
is empty (no medium inside the cavity).

\begin{figure}[!tbp]
\rput(7,4.5){a)}
\includegraphics[width=.95\linewidth]{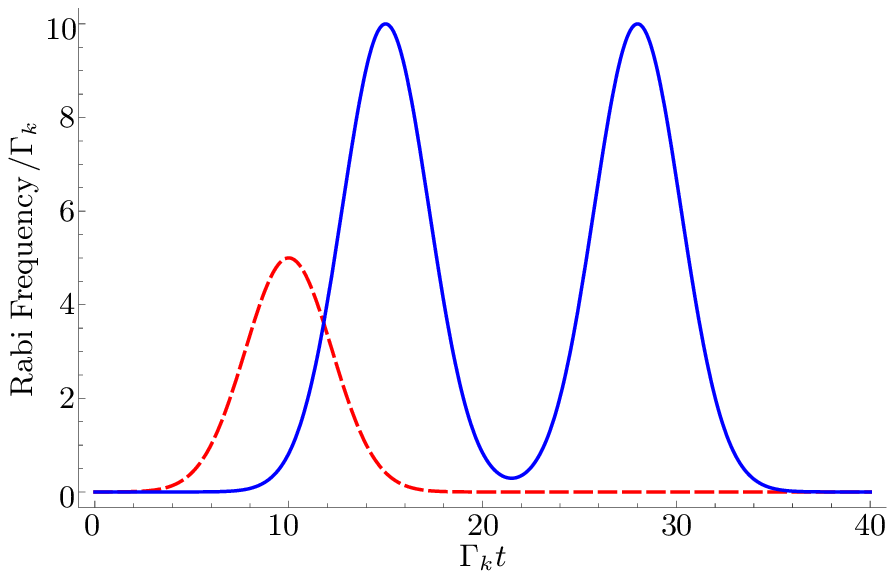}
\\[2ex]
\rput(7,4.5){b)}
\includegraphics[width=.95\linewidth]{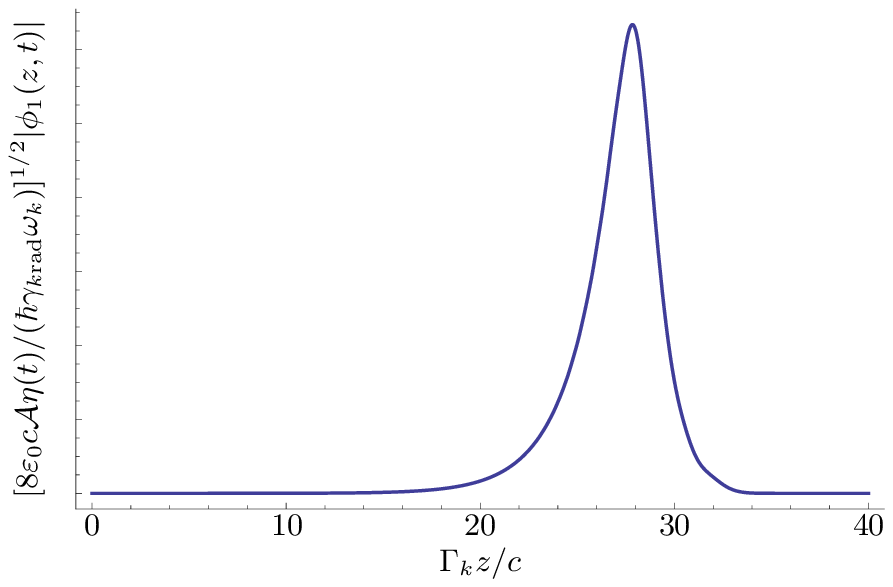}
\caption{\label{fig5.3}
(Color online)
a) Sequence of the interactions of the atom with the cavity field
[$R_k g_c(t)$, dashed curve]
and
with the pump field
[$\Omega _p(t)$, solid curve].
b) Numerical result for the spatio-temporal
shape of the excited outgoing mode for $\Gamma_k t$ $\!=$ $\!40$.
}
\end{figure}

The behavior of the absolute value of
the spatio-temporal shape $\phi_1(z,t)$
of the excited outgoing wave packet is illustrated in
Figs.~\ref{fig5.1} and \ref{fig5.3},
where the maximum
single-mode vacuum Rabi frequency of the
atom--cavity-field interaction
is assumed to be $R_k$ $\!=$
$\!\sqrt{\alpha_k\omega_k}$ $\!=$ $\!5$ $\!\times$ $\!\Gamma_k$,
and $(\Omega_p$,
$\!\Delta_p$, $\!\omega_k - \omega_0)$ $\!=$ $(\!10$,
$\!10^{-3}$, $\!10^{-3})$ $\!\times$ $\!\Gamma_k$.
The (numerical) results show that the spatio-temporal
shape of the outgoing wave packet
is determined by the cavity decay rate as well as the
pump process, in particular by the delay between
the interaction of the atom with the pump field and
the cavity-assisted field,
$\Omega_p(t)$ and $g_c(t)$ respectively.
The efficiency of one-photon Fock state preparation $\eta(t)$ is
an almost monotonically increasing function that
reaches a constant value equal to the ratio of
the cavity radiative decay rate to the total
decay rate, which is assumed to be 90\% in the following.
Note that the cavity total decay rate
is the sum of the decay rates due to radiative
input--output coupling and due to unwanted losses such as scattering
and medium absorption.
In the same way, a sequence of light pulses
each
of almost
one-photon Fock
state
can be generated by means of a beam of atoms that cross the cavity one at
a time or by the same atom introducing recycling field pulses,
which bring the atomic population to the initial level $|1\rangle$.
It should be pointed out that
in the case of stimulated Raman adiabatic passage
the spatio-temporal
shape of the excited outgoing wave packet
is independent of the shape of
the pump field, as can be seen from a comparison of
Fig.~\ref{fig5.1} with Fig.~\ref{fig5.3}.

\begin{figure}[!tbp]
\rput(7,4.5){a)}
\includegraphics[width=.95\linewidth]{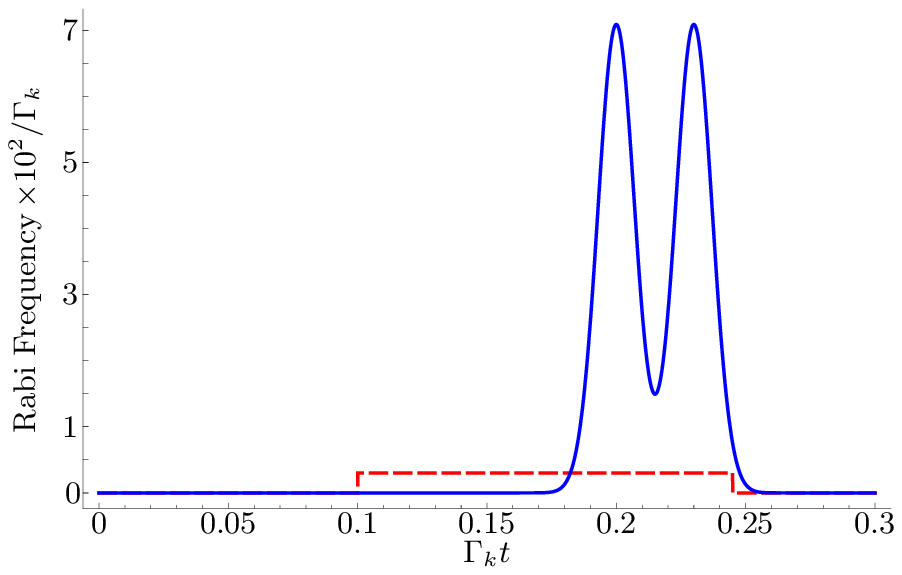}
\\[2ex]
\rput(7,4.5){b)}
\includegraphics[width=.95\linewidth]{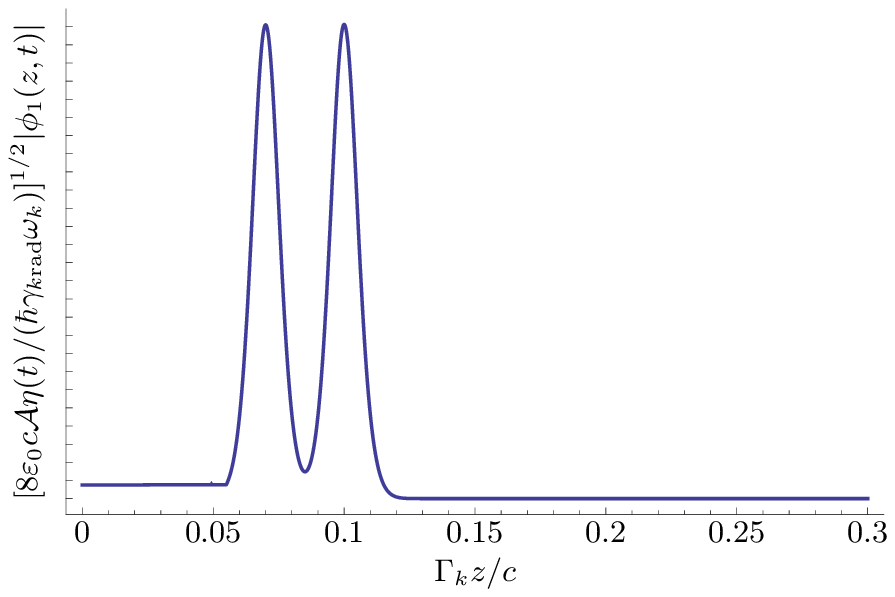}
\caption{\label{fig5.4}
(Color online)
a) Sequence of the interactions of the atom with the cavity field
[$R_k g_c(t)$, dashed curve]
and
with the pump field
[$\Omega _p(t)$, solid curve].
b) Numerical result for the spatio-temporal
shape of the excited outgoing mode for $\Gamma_k t$ $\!=$ $\!0.3$.
}
\end{figure}
\begin{figure}[!tbp]
\rput(7,4.5){a)}
\includegraphics[width=.95\linewidth]{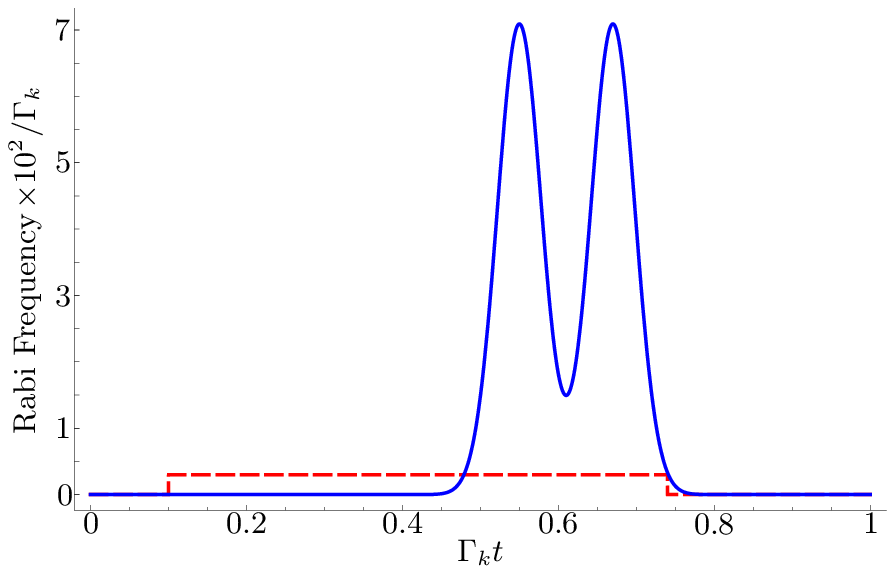}
\\[2ex]
\rput(7,4.5){b)}
\includegraphics[width=.95\linewidth]{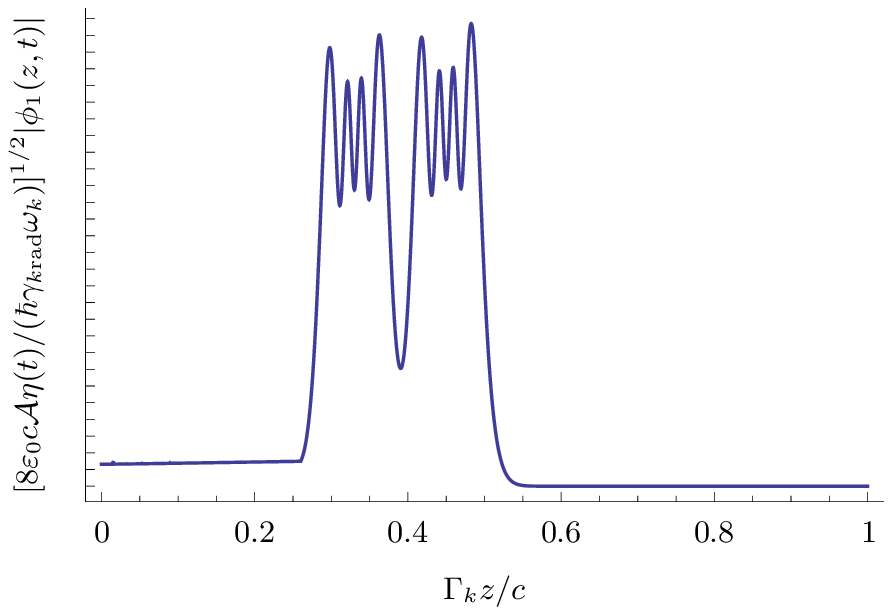}
\caption{\label{fig5.5}
(Color online)
a) Sequence of the interactions of the atom with the cavity field
[$R_k g_c(t)$, dashed curve]
and
with the pump field
[$\Omega _p(t)$, solid curve].
b) Numerical result for the spatio-temporal
shape of the excited outgoing mode for $\Gamma_k t$ $\!=$ $\!1$.
}
\end{figure}
\begin{figure}[!tbp]
\rput(7,4.5){a)}
\includegraphics[width=.85\linewidth]{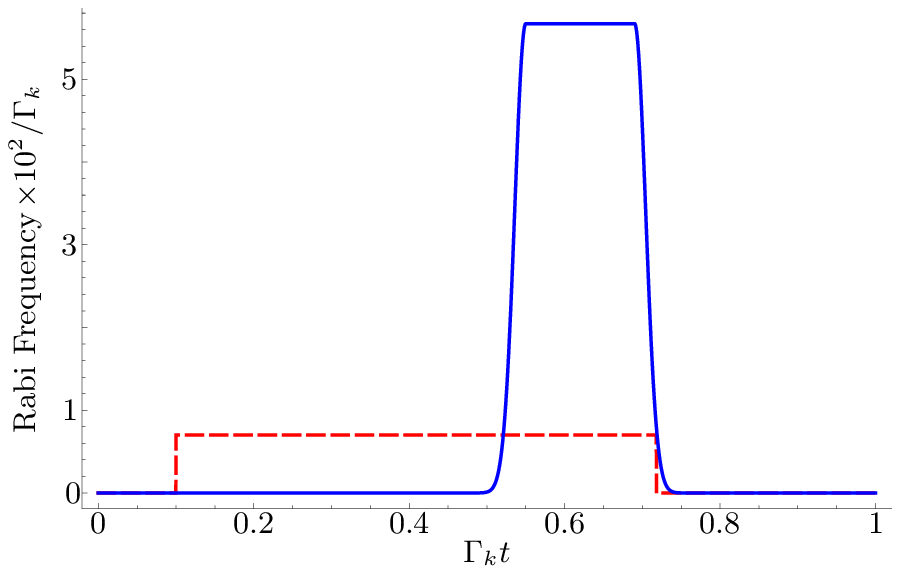}
\\[2ex]
\rput(7,4.5){b)}
\includegraphics[width=.8\linewidth]{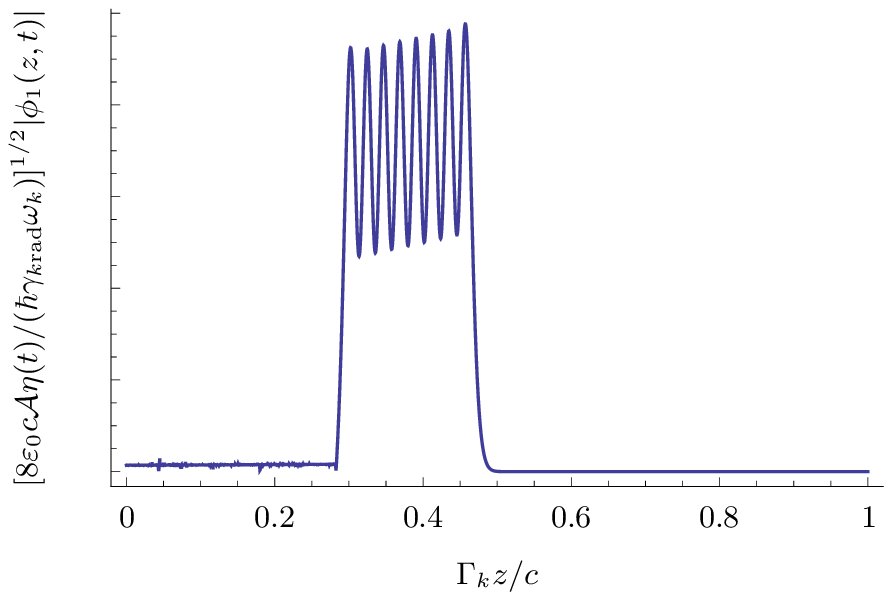}
\caption{\label{fig5.8}
(Color online)
a) Sequence of the interactions of the atom with the cavity field
[$R_k g_c(t)$, dashed curve]
and
with the pump field
[$\Omega _p(t)$, solid curve].
b) Numerical result for the spatio-temporal
shape of the excited outgoing mode for $\Gamma_k t$ $\!=$ $\!1$.
}
\end{figure}

The interaction of a classically pumped single three-level atom
in a cavity also allows the generation of
an outgoing wave packet
carrying a one-photon Fock state with
given spatio-temporal shape,
adjusted by means of the pump field shape.
The control of
the spatio-temporal shape of the outgoing
wave packet can be achieved when the pump field
is intense enough and very short
compared to the cavity decay time.
As illustrated in \mbox{Figs.~\ref{fig5.4}--\ref{fig5.8}}, let
us assume
that the interaction of the atom with the cavity field
can be regarded as being constant on
this time scale and is turned off
before the pump field is turned off.
For a twin-peaked pump field shape and $(R_k$, $\Omega_p$,
$\!\Delta_p$, $\!\omega_k - \omega_0)$ $\!=$ $(\!30$, $\!700$,
$\!0.001$, $\!0.001)$ $\!\times$ $\!\Gamma_k$,
Fig.~\ref{fig5.4} reveals that
the leading edge of the spatio-temporal shape of the
excited outgoing wave packet precisely reproduces
the shape of the pump field.
However, since the regime of
stimulated adiabatic Raman passage is
not realized,
the efficiency of one-photon Fock state preparation is
low \mbox{$\eta(t\rightarrow \infty)$ $\!=$ $\!10^{-4}$}.

To achieve higher values of the efficiency, the duration
and/or the intensity of the pump pulse
should be increased.
As it is illustrated in
Figs.~\ref{fig5.5} and~\ref{fig5.8} for
$(R_k$, $\Omega_p$,
$\!\Delta_p$, $\!\omega_k - \omega_0)$ $\!=$ $(\!30$, $\!700$,
$\!0.001$, $\!0.001)$ $\!\times$ $\!\Gamma_k$
and $(R_k$, $\Omega_p$,
$\!\Delta_p$, $\!\omega_k - \omega_0)$ $\!=$ $(\!70$, $\!570$,
$\!0.001$, $\!0.001)$ $\!\times$ $\!\Gamma_k$, respectively,
the spatio-temporal shape of the excited outgoing
wave packet features Rabi oscillations
according to
the interaction of the atom with the pump pulse
and follows the shape of the pump pulse.
For the twin-peaked pulse in Fig.~\ref{fig5.5} and
the square-like pulse in Fig.~\ref{fig5.8}, respectively,
efficiencies of
$\eta(t\rightarrow \infty)$ $\!=$ $\!0.014$
and $\eta(t\rightarrow \infty)$ $\!=$ $\!0.022$
are observed.
Note that with increasing intensity of the pump field, the
oscillations of the spatio-temporal shape
become very
fast
and
may be
therefore
not resolvable.

To conclude, we have shown that the interaction of a
pumped three-level
$\Lambda$-type atom with the electromagnetic field in a cavity can
be used to generate one-photon Fock states in
well-defined outgoing wave packets.
In particular in the case when
stimulated Raman adiabatic passage is realized, then the efficiency of
one-photon Fock state generation
may be close to $100\%$; it is only limited
by the unavoidable unwanted
losses in the system such scattering and medium absorption.
In particular, for large times the efficiency equals to
the ratio of the cavity radiative decay rate to the total decay
rate. The spatio-temporal shape
of the outgoing wave packet
is robust against fluctuations
of the shape of the atom--pump interaction, and,
therefore, the scheme can be used to produce
identical wave packets on demand carrying one-photon Fock
states each.
For sufficiently intense and short pumping, i.\,e.,
beyond the adiabatic regime, the scheme enables
control of
the spatio-temporal shape of the excited outgoing
wave packet
by means of the atom--pump interaction shape. However
in this case, the
efficiency of one-photon Fock state generation is
rather low in general.

\begin{acknowledgments}
This work was supported by the Deutsche Forschungsgemeinschaft.
\end{acknowledgments}


\bibliography{bibl}

\end{document}